\documentclass[a4paper, 11pt]{article}

\usepackage{tex_andy}
\usepackage{algorithms_andy}
\usepackage{theorems_andy}
\usepackage{equations_andy}

\usepackage{k_tex}
\usepackage{k_figures}
\usepackage{misc_k,Labels}

\usepackage{graphicx,epic,eepic}
\usepackage{epsfig}

\makeatletter
\@ndyTheorem{Invar}{Invariance}{Invariance}
\setcmd{\bInvar}[1]{\begin{Invar}\Label{Invar:#1}}
\setcmd{\eInvar}{\end{Invar}}
\setcmd{\rInvar}[1]{Invariance \Ref{Invar:#1}}
\makeatother
\setcommand{\Intvl}[2]{\:[#1, #2)\:}

\usepackage{cite}
\usepackage{url}

\begin{document}
  \setcmd{\ltu}{Lule{\aa} University of Technology, Sweden}
\setcmd{\upr}{University of Primorska, Slovenia}
\setcmd{\imfm}{Institute of Mathematics, Physics and Mechanics, Slovenia}
\setcmd{\uw}{School of Computer Science, University of Waterloo,
  Ontario, Canada}
\setcmd{\andyMail}{andrej.brodnik@upr.si}
\setcmd{\kMail}{johan.karlsson@csee.ltu.se}
\setcmd{\ianMail}{imunro@uwaterloo.ca}
\setcmd{\andreasMail}{andreas.nilsson@csee.ltu.se}

\setcmd{\papertitle}{An O(1) Solution to the Prefix Sum Problem on a
  Specialized Memory Architecture}

\author{
  Andrej Brodnik \thanks{\ltu} \thanks{\upr} \thanks{\imfm}
\and
  Johan Karlsson
  $^\ast$
\and
  J.~Ian Munro
  \thanks{\uw}
\and
  Andreas Nilsson
  $^\ast$
\and
\\
     {\andyMail \ \ \ \kMail}
\and {\ianMail \ \ \ \andreasMail}
}

\date{}

\title{\papertitle}

\maketitle

\begin{abstract}
  In this paper we study the Prefix Sum problem introduced by
Fredman.
We show that it is possible to perform both update and
retrieval in $O(1)$ time simultaneously under a memory model in which
individual bits may be shared by several words.
We also show that two variants (generalizations) of the problem can be
solved optimally in $\Theta(\lg N)$ time under the comparison based
model of computation.


\end{abstract}

\bSec{Introduction}{intro}
  In this paper we discuss solutions to variants of the \emph{Prefix Sum
problem} (i.e. finding the sum of the first $j$ elements in an array and
also updating these values) which was introduced by Fredman
\cite{JACM::Fredman:1982:CMACPS}.
Various lower bounds have been proven for the problem. We, however,
focus on the problem under a nonstandard, though very feasible, model
to achieve a constant time solution.
In particular, we focus primarily on the so called \emph{RAMBO} model of
computation, which is an extension of the random access machine (RAM),
that is a Random Access Machine with Byte Overlap, i.e. a bit can be
in several words.  This model was first suggested by Fredman and Saks
\cite{STOC::FredmanS:1989:CPCDDS} and further described and used by
Brodnik~\etal\cite{JSS::BrodnikCFKM:2005:WCCTPQ,PhD:Brodnik:1995}.

Fredman and Saks actually suggested the RAMBO model in connection with
the Prefix Sum problem. They claim, with no hint of how it may be
done, that Prefix Sum mod 2 can be solved in constant time under the
model.  We show how this can be done not only for Prefix Sum mod 2 but
for Prefix Sum modulo an arbitrary universe size $M \leq
2^{\Theta(b/n)}$.

The RAMBO model, besides the usual RAM operations (\cf
\cite{HTCS:Boas:MMS}), also has a part of memory where a bit may occur
in several registers or in several positions in one register. The way
the bits occur in this part of the memory has to be specified as part
of the model.
One example of such a memory variant is a square of bits with $n$ rows
and $n$ columns. A $n$-bit word can be fetched either as a row or a
column. In such a memory each bit can be accessed either by the row
word or the column word.

Brodnik~\etal\cite{JSS::BrodnikCFKM:2005:WCCTPQ} use a variant of
RAMBO, referred to as the Yggdrasil variant, to solve the Priority
Queue problem in $O(1)$ worst case time. That variant has been
implemented in hardware\cite{ECSC::BrodnikKLMST:1999:DHPMMPC100} and
the actual rerouting of the bits on a word fetch is not difficult.
In this paper we modify the Yggdrasil variant slightly and solve the
Prefix Sum problem. This gives further evidence of the value of such
an architecture, at least for a special purpose processor.

Now let us formally define the Prefix Sum problem:
\bDefn{PSP}
The \emph{Prefix Sum problem} is to maintain an array,
$\mathcal{A}$, of size $N$, and to support the following
operations:\\
\begin{tabular}{lp{5.5cm}}
\Code{Update($j$, $\Delta$)}&$\mathcal{A}(j) := \mathcal{A}(j) + \Delta$\\
\Code{Retrieve($j$)}&return $\sum_{i=0}^{j}{\mathcal{A}(i)}$
\end{tabular}\\
where $0 \leq j < N$.
\eDefn
Fredman showed that, under the comparison based model of computation,
an $O(\lg N)$ solution exists for the Prefix Sum
problem\cite{JACM::Fredman:1982:CMACPS}.

The problem can be generalized in several ways and we start by adding
another parameter, $k$ to the \Code{Retrieve} operation. This
parameter is used to tell the starting point of the array interval to
sum over. Hence, \Code{Retrieve(k,j)} returns
$\sum_{i=k}^{j}{\mathcal{A}(i)}$, where $0 \leq k \leq j < N$. This
variant is usually referred to as the \emph{Partial Sum} or
\emph{Range Sum problem}. The Partial Sum problem can be solved using
a solution to the Prefix Sum problem (\Code{Retrieve(k,j) = }
\Code{Retrieve(j) - Retrieve(k-1)}). In fact, the two problems are
often used interchangeably.

Furthermore, there is no obvious reason to only allow addition in the
\Code{Update} and \Code{Retrieve} operations. We can allow any binary
function, $\oplus$, to be used. In fact we can allow the \Code{Update}
operation to use one function, $\oplus_u$, and the \Code{Retrieve}
operation to use another function, $\oplus_r$.
%
We will refer to this variant of the problem as the \emph{General
  Prefix Sum problem}.

Moreover, one can allow array position to be inserted at or deleted
from arbitrary places. Hence, we can have sparse arrays, \eg an array
where only $\mathcal{A}(5)$ and $\mathcal{A}(500)$ are present.
Positions which have not yet been added or have been deleted have the
value $0$. We refer to this variant as the \emph{Dynamic Prefix Sum
  problem}. Brodnik and Nilsson\cite[pp 65-80]{Lic:Nilsson:2004}
describe a data structure they call a BinSeT tree which can be
modified slightly to support all operation of the Dynamic Prefix Sum
problem in $O(\lg N)$ time.
Another generalization is to use multidimensional arrays and this
variant has been studied by the data base
community\cite{DAVAK::BengtssonC:2004:SERSQOLAP, ho97, geffner99,
  geffner99j, riedewald00, riedewald01}.

Several lower bounds have been presented for this problem: Fredman
showed a $\Omega(\lg N)$ algebraic complexity lower bound and a
$\Omega(\lg N/\lg \lg N)$ information-theoretic lower
bound\cite{JACM::Fredman:1982:CMACPS}.
Yao\cite{SIAMJoC:Yao:1985:CMPS} has shown that $\Omega(\lg N/\lg \lg
N)$ is an inherent lower bound under the semi-group model of
computation and this was improved by Hampapuram and Fredman
to $\Omega(\lg N)$\cite{SIAMJoC::HampapuramF:1998:OBBTCMPS}.
We side step these lower bound by considering the RAMBO model of
computation.

As with all RAM based model we need to restrict the size
of a word which can be stored and operated on. We denote the word size
with $b$ and assume that $b=2^{O(1)}$ which is true for most computers
today.
A bounded word size also implies a bounded universe of elements that
we store in the array. We use $M$ to denote the universe size.
Hence all operations $\oplus$ have to be computed modulo $M$ and we
require that each of the operands and the result are stored in one
word.

We will use $n$ and $m$ to denote $\lceil \lg N \rceil$  and $\lceil \lg M
\rceil$ respectively. Hence, $N \leq 2^n$ and $M \leq 2^m$.
Both $n$ and $m$ are less than or equal to $b$, ($n,m \leq b$). In one
of the solutions we actually require that $nm \leq b$.

In \sectref{psp} we show a $O(1)$ solution to the Prefix Sum problem
under the RAMBO model using a modified Yggdrasil variant. In
\sectref{gpsp} we discuss a $O(\lg N)$ solution to the General and
Dynamic Prefix Sum problems and finally conclude the paper with some
open questions in \sectref{concl}.


\bSec{An $O(1)$ Solution to the Prefix Sum Problem}{psp}
  In our O(1) solution to the Prefix Sum problem we use a
complete binary tree on top of the array (\figref{tree}).
We label the nodes in standard heap order, \iE, the root is node
$\nu_1$ and the left and right children of a node $\nu_i$ are
$\nu_{2i}$ and $\nu_{2i+1}$ respectively.
In each node we store $m$ bits representing the sum of the
leaves in the left subtree.
Since we build a complete binary tree on top of the array we assume
that $N = 2^n$ (if this is not true we still build the complete tree
and in worst case waste space proportional to $N/2-1$).
We do not store the original array $\mathcal{A}$ since its values are
stored implicitly in the tree. The only value not stored in the tree
(if $N=2^n$ only) is $\mathcal{A}(N-1)$ and we store this value
explicitly (\Code{vn1}).
Formally we define:
\bDefn{m-tree}
A \emph{N-m-tree} is a complete binary tree with $N$ leaves in which
the internal nodes ($\nu_i$) store a $m$-bit value. In addition, a
$m$-bit value is stored separately (\Code{vn1}).
\eDefn

To update $\mathcal{A}(j)$ (\rAlg{update-n}) in this structure we
have to update all the nodes on the path from leaf $j$ to the root in
which $j$ belongs to the left subtree. To \Code{Retrieve($j$)}
(\rAlg{retrieve-n}) we need
to sum the values of all the nodes on the path from leaf $j+1$ to the
root in which $j+1$ belongs to right subtree.
Note that the path corresponding to array position $j$ starts at node
$\nu_{N/2 + j/2}$.

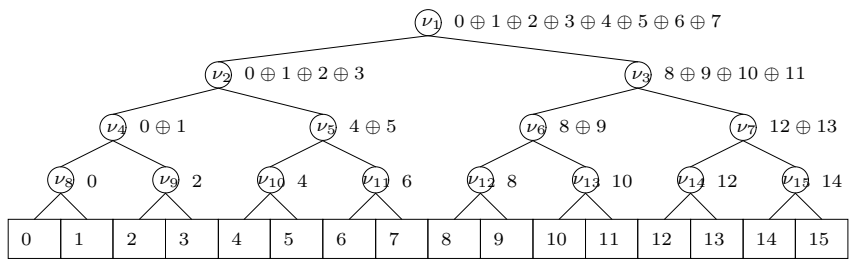
\figinput{tree}{tree.eepic}{Complete binary tree ontop of
  $\mathcal{A}$. Nodes are storing the sum of the
  values in the leaves covered by the left subtree.}

\bAlg{update-n}{pascal}{Updating of a N-m-tree in $O(\lg N)$ time.}
update($j$, $\Delta$)
  if (j == N-1)
    vn1 = vn1 + $\Delta$;
  else
    i = N + j;
    while (i > 1)
      next = $i \div 2$;
      if ($i \bmod 2$ == 0)
        $\nu_{next}$ = $\nu_{next}$ + $\Delta$ $\bmod M$);
      i = next;
\eAlg

\bAlg{retrieve-n}{pascal}{Retrieve in a N-m-tree in $O(\lg N)$ time.}
retrieve($j$)
  if (j == N-1)
    sum = vn1;
    i = N + j;
  else
    sum = 0;
    i = N + j + 1;

  while (i > 1)
    next = $i \div 2$;
    if ($i \bmod 2$ == 1)
      sum = sum + $\nu_{next}$ $\bmod M$;
    i = next;

  return sum;
\eAlg

The method described above implies a $O(\lg N)$ update and retrieval
time in the RAM model.
To achieve constant time update and retrieval we use a variant of the
RAMBO model similar to the Yggdrasil variant. In the Yggdrasil
variant, registers overlap as paths from leaf to root in a complete
binary tree with one bit stored in each internal
node\cite{JSS::BrodnikCFKM:2005:WCCTPQ}. We generalize the Yggdrasil
variant and let it store $m$ bits in each node and call this variant
\emph{m-Yggdrasil}.
In any m-Yggdrasil, register \Code{reg[i]} corresponds to the path
from node $\nu_{N/2 + i}$ to the root of the tree. Each register
consists of $nm \leq b$ bits. In total the m-Yggdrasil registers
need $(N-1) \cdot m$ bits.

Now, we use the registers from m-Yggdrasil to store the nodes of our
tree. The path corresponding to array position $j$ is stored in
\Code{reg[j/2]} and hence all nodes along the path can be accessed
at once.

We let levels of the tree be counted from the internal nodes above the
leaves starting at $0$ and ending with $n-1$ at the root. If the $i$th
bit of $j$ is $1$ then $j$ is in the right subtree of the node on
level $i$ of the path and in the left otherwise. Hence $j$ can be used
to determine which nodes along the path should be updated (nodes
corresponding to bits of $j$ that are $0$) and which nodes should be
used when retrieving a sum (nodes corresponding to bits of $j$ that
are $1$).

When updating the m-Yggdrasil registers (\rAlg{update-n-reg}), for all
bits of $j$, if the $i$th bit of $j$ is $0$ we add $\Delta$ to the
value of the $i$th node along the path from $j$ to the root. To do
this we shift $\Delta$ to the corresponding position ($\Delta <<
(im)$) and add to \Code{reg[j/2]}.
Instead of checking whether the $i$th bit of $j$ is $0$ we can mask the
shifted $\Delta$ with a value based on $\NOT j$. The value consists
of, if the $i$th bit of $\NOT j$ is $1$,  $m$ $1$s shifted to the
correct position and $m$ $0$s otherwise.

\bAlg{update-n-reg}{pascal}{Updating of a N-m-tree stored in
  m-Yggdrasil memory ($O(\lg N)$ time).}
update($j$, $\Delta$)
  if (j == N-1)
    vn1 = vn1 + $\Delta$;
  else
    for (i=0; 0 < n; i++)
      if (((j >> i) $\AND$ 1) == 0)
        reg[j/2] = reg[j/2] + ($\Delta$ << (i*m));
\eAlg

Actually, as long as the binary operation only affects the $m$
bits that should be updated we can use word-size parallelism (\cf
\cite{PhD:Brodnik:1995}) and perform the update of all nodes
in parallel. In \rSec{oplus+} we show that addition modulo
$M$ can be implemented affecting only $m$ bits.

We use two functions (\Code{dist(i)} and \Code{mask(i)}) to simplify the
description of the update and retrieve methods.
The function \Code{dist(i)}, ($0 \leq i < 2^{m}$) computes $nm$-bit
values. The values are $n$ copies of the $m$ bits in
$i$.  For example, given $m=3, n=4$ \Code{dist(010)} is
$010010010010$.
The function \Code{mask(i)}, ($0 \leq i < 2^{n}$) also computes $nm$-bit
values. These values are computed as follow: bit $j$ ($0 \leq j <
n$) of $i$ is copied to bits $jm..(j+1)m-1$.  For example, given $m=3,
n=4$, \Code{mask(1001)} is $111000000111$.
Both these functions can be implemented by using word-size
parallelism\cite{PhD:Brodnik:1995}.

We can update the tree in constant time using the procedure in
\rAlg{update-reg}. First we make $n$ copies of $\Delta$ and then mask
out the copies we need. Then finally we add this to \Code{reg[j/2]}
and the masked distributed $\Delta$ and store the result in
\Code{reg[j/2]}. For the case when $j=N-1$ we simply add to \Code{vn1}
and $\Delta$ and store it in \Code{vn1}. This gives us the following
lemma:
\bLemma{update}
The update operation of the Prefix Sum problem can be supported
in $O(1)$ when parts of the N-m-tree is stored in a m-Yggdrasil memory.
\eLemma

\bAlg{update-reg}{pascal}{Updating of a N-m-tree stored in m-Yggdrasil
  memory using word size parallelism ($O(1)$ time).}
update($j$, $\Delta$)
  if (j == N-1)
    vn1 = vn1 + $\Delta$;
  else
    reg[j/2] = reg[j/2] + (dist($\Delta$) $\AND$ mask($\NOT$ j));
\eAlg

To support the retrieve method in constant time we use a table
\Code{SUM[i]}, ($0 \leq i < 2^{nm}$) with $m$-bit values that are the
sum modulo $M$ of the $n$ $m$-bit values in $i$.

To retrieve the sum (\rAlg{retrieve-reg}) we read the register
\Code{reg} corresponding to $j$ and mask out the parts we need. Then
we use the table \Code{SUM} to calculate the sum. Finally, we add
\Code{vn1} to the sum if $j=N-1$.

\bAlg{retrieve-reg}{pascal}{Retrieve in a N-m-tree stored in
  m-Yggdrasil memory using word size parallelism ($O(1)$ time).}
retrieve($j$)
  if (j == N-1)
    v = reg[j/2] $\AND$ mask(j);
  else
    v = reg[(j+1)/2] $\AND$ mask(j+1);

  sum = SUM[v];

  if (j == N-1)
    sum = vn1 + sum;

  return sum;
\eAlg

The space needed by the table \Code{SUM} is $2^{nm}\cdot m =
N^{\lg{M}} \cdot m = M^{\lg{N}} \cdot m$, which is rather large. In
order to reduce the space requirement we can reduce, by half, the
number of bits used as index into the table. This gives us a space
requirement of $\sqrt{M^{\lg{N}}} \cdot m$. We do this by shifting
the top $n/2$ $m$-bit values from \Code{reg} down and computing the
sum modulo $M$ of these values and the bottom $n/2$
values. Then this new $(n/2)m$-bit value is used as index into
\Code{SUM} instead.

We can actually repeat this process until we get the $m$-bit we
desire, and hence we do not need the table \Code{SUM}
(\rAlg{retrieve-reg-lg}). However, this does increase the time
complexity to $O(\lg{n}) = O(\lg\lg{N})$.
This gives us a trade off between space and time. By allowing
$O(\iota)$ steps for the retrieve method we need $M^{\lg{N}/2^\iota}
\cdot m$ bits for the table.

\bAlg{retrieve-reg-lg}{pascal}{Retrieve in a N-m-tree stored in
  m-Yggdrasil memory using no additional memory ($O(\lg \lg N)$ time).}
retrieve($j$)
  if (j == N-1)
    v = reg[j/2] $\AND$ mask(j);
  else
    v = reg[(j+1)/2] $\AND$ mask(j+1);

  $\iota$ = $\lceil\lg{n}\rceil$;
  do
    $\iota$ = $\iota$-1;
    vnew = (v>>(($2^\iota$)m)) + (v $\AND$ ((1<<(($2^\iota$)m))-1));
    v = vnew;
  while ($\iota$ > 0)

  if (j == N-1)
    sum = vn1 + sum;

  return sum;
\eAlg

\bLemma{retrieve}
The retrieve operation of the Prefix Sum problem can be
supported in $O(\iota + 1)$ time using $O(M^{\lg{N}/2^\iota} \cdot m +
m)$ bits of memory in additions to the N-m-tree. Parts of the N-m-tree
is stored in m-Yggdrasil memory.  \eLemma
By adjusting $\iota$ we can achieve the following result:
\bCoro{retrieve-1-lgN}
The retrieve operation of the Prefix Sum problem can be supported
in:
\begin{itemize}
\item $O(1)$ time using $O(M^{(\lceil\lg N\rceil)/2} \cdot m)$ bits of memory
  in additions to the N-m-tree, with $\iota=1$.
\item $O(\lg \lg N)$ time using $O(m)$ bits of memory
  in additions to the N-m-tree, with $\iota=\lceil \lg \lg N \rceil$.
\end{itemize}
\eCoro

\bSSec{Addition modulo $M$}{oplus+}
Let us consider the two $m$-bit operands $a$ and $b$ which are split
into two pieces each ($a_{lo}, a_{hi}, b_{lo}$ and $b_{hi}$). The two
pieces $a_{lo}$ and $a_{hi}$ contain the $m/2$ least and most
significant bits of $a$ respectively (similarly for $b_{lo}$ and
$b_{hi}$). Note that $a_{lo}$ and the other pieces are stored in
$m$-bit but only the $m/2$ least significant bits are used.

We can now add the the two operands
\bEquArr{c}
c1_{lo} &=& a_{lo} + b_{lo}\\
c1_{hi} &=& a_{hi} + b_{hi}\enspace.
\eEquArr
However, both $c1_{lo}$ and $c1_{hi}$ might need $m/2+1$ bits for its
result. The $m/2+1$ bit of $c1_{lo}$ should be added to $c1_{hi}$ and we
split $c1_{lo}$ into two pieces ($c1_{lo,lo}$ and $c1_{lo,hi}$) and add
the most significant bits to $c1_{hi}$,
\bEquArr{c2}
c_{hi} &=& c_{hi} + c_{lo,hi}\\
c_{lo} &=& c_{lo,lo}\enspace.
\eEquArr
The result of $a+b$ is now stored in $c_{lo}$ and $c_{hi}$
and we have not used more than $m$ bits in any word. However, in total
$m+1$ might be needed for the value.

To compute $c \bmod M$ we can check whether or not $c - M >= 0$, if so
$c \bmod M = c - M$ and otherwise $c \bmod M = c$.  However, we do not
want to produce a negative value since that would affect all the bits
in the word. Instead we add an additional $2^m$ to the value and
compare to $2^m$, \ie $c + 2^m - M \geq 2^m$. Since $2^m - M \geq 0$
this will never produce a negative value. Note that $c + 2^m - M < M
- 1 + M - 1 + 2^m - M = M + 2^m - 2 <= 2^{m+1} - 2$ which only needs
$m+1$ to be represented. Hence, if we calculate this value using the
strategy above we will not use more than $m$ bits of any word.

Furthermore, a straight forward less than comparison can not be
performed using word-size parallelism since all bits of the words are
considered. Instead we view the comparison as a check whether the
$m+1$st bit is set or not. If it is set the value is larger than or
equal to $2^m$.
We can actually create a bit mask which consists of $m$ $1$s if
the $m+1$st bit is set and $m$ $0$s otherwise
\bEquArr{c3}
  d = (c+2^m-M \AND 2^{m}) - ((c+2^m-M \AND 2^{m}) >> m)\enspace.
\eEquArr
This bit mask $d$ can then be used to calculate $res = c \bmod M$.
Since $res$ is equal to $c-M$ if the $m+1$st bit of $c$ is set
and $c$ otherwise we get
\bEquArr{c4}
  res = ((c-M) \AND d) \OR (c \AND \NOT d)\enspace.
\eEquArr
When computing $c-M$ we must make sure that we do not produce a
negative value. This is done by using a similar strategy as for
addition above, but we also set any of the bits in $c_{hi,hi}$ to $1$
during the computation. If $c-M$ is greater than $0$ this will not
affect the result and otherwise the result will not be used.

We have a procedure which can be used to compute $(a+b) \bmod M$
without using more than $m$ bits in any word. Hence, word-size
parallelism can be used and we get our main result from this section:
\bTheo{AMP-oplus-ygg}
Using the N-m-tree together with the m-Yggdrasil memory we can support
the operations of the Prefix Sum problem in $O(\iota + 1)$ time
using $(N-1)m$ bits of m-Yggdrasil memory and $O(M^{n/2^\iota} \cdot m
+ m)$ bits of ordinary memory.
\eTheo


\bSec{An $O(\lg N)$ Solution to the General and Dynamic Prefix Sum
  Problem}{gpsp}
  We can actually partially solve the General Prefix Sum problem using
the N-m-tree data structure and the m-Yggdrasil variant of RAMBO.  All
binary operations such that all elements in the universe have a unique
inverse element (\ie binary operations which form a \emph{Group} with
the set of elements in the universe) and only affect the $m$ bits
involved in the operation can be supported. This includes for example
addition and subtraction but not the maximum function.

To solve the General and Dynamic Prefix Sum problem for semi-group
operations we modify the Binary Segment Tree (BinSeT) data structure
suggested by Brodnik and Nilsson.  It was designed to handle
in-advance resource reservation\cite[pp 65-80]{Lic:Nilsson:2004} and
if it is slightly modified it can solve both the General and Dynamic
Prefix Sum problems efficiently.
The original BinSeT stores, in each internal node, $\mu$, the
maximum value over the interval, and $\delta$, the change of the value
over the interval. Further, it also stores $\tau$, the time of the
left most event in the right subtree.

Instead of storing times as interval dividers we store array indices.
To solve the Dynamic Prefix Sum problem with addition as operation and
we only need to store $\delta$. When solving the General and Prefix Sum
problem one need to store information depending on the two binary
operations $\oplus_u$ and $\oplus_r$.

When adding a new array position or deleting an array position the
tree is rebalanced
(cf.~\cite{::Adelson-VelskiiL:1962:AOI,Levitin:2003:IDAA}) and hence
the height is always $O(\lg N)$.
When updating a value in an array position we start at the root and
search for the proper leaf using the interval dividers. During the
back tracking of the recursion we update the information stored in
each affected node.

At retrieval we process the information of the proper nodes when
traversing the tree.
Since the height of the tree is $O(\lg N)$ all the operations can be
performed in $O(\lg N)$ time. This matches the lower bound by
Hampapuram and Fredman\cite{SIAMJoC::HampapuramF:1998:OBBTCMPS}

BinSeT consists of $O(N)$ nodes when we use it to solve the General
Prefix Sum. Each node contains $O(1)$ $m$-bit values and hence the
total space requirement is $O(Nm)$ bits.


\bSec{Conclusion}{concl}
  The Dynamic and General Prefix Sum problems can both be solved
optimally in $\Theta(\lg N)$ using $O(Nm)$ space under the comparison
based model with semi-group operations.

The Prefix Sum problem can be solved in $O(1)$ time under the
RAMBO model when we allow $O(\sqrt{M^{(\lceil\lg N\rceil)}} \cdot m)$
bits of ordinary memory and $O(Nm)$ bits of $m$-Yggdrasil memory to
be used. This is a huge amount of ordinary memory and if we restrict
the space requirement to be sub exponential in both $N$ and $M$ ($O(m)$
bits of ordinary memory and $O(Nm)$ bits of $m$-Yggdrasil memory) we
need to used $O(\lg \lg N)$ time.
We know of no better lower bound under RAMBO than the trivial
$\Omega(1)$ when only allowing $O((N^{O(1)}+M^{O(1)})m)$ space.

Further, it is currently unknown if one can achieve a $O(1)$ solution
to the Dynamic and General Prefix Sum problems using the RAMBO model.
Another open question is whether or not it is possible achieve a
$o(\lg N)$ solution to the multidimensional variant.

\bibliography{strings,misc,books,theses,copies,publications,proceedings,other}
\bibliographystyle{plain}


\end{document}